\documentclass[prd,showpacs,showkeys,floatfix,nofootinbib,eqsecnum,
                 preprint,12pt,tightenlines,fleqn]{revtex4} 

\usepackage{amsmath,amssymb,revsymb,graphicx,dcolumn}

\newcommand  {\version}{v7}            

\newcommand{\beq}{\begin{equation}}
\newcommand{\eeq}{\end{equation}}
\newcommand{\beqa}{\begin{eqnarray}}
\newcommand{\eeqa}{\end{eqnarray}}
\newcommand{\diag}{\ensuremath{\mathrm{diag}}}  
\newcommand{\GammaVM}{\Gamma_\text{VM}}         
\newcommand{\gammaV}{\gamma}                    

\begin{document}

\noindent Phys. Rev. D 78, 083533 (2008) \hfill arXiv:0803.0281 [gr-qc] (\version)\vspace*{1\baselineskip}
\title{Equilibrium boundary conditions, dynamic vacuum energy,\\
       and the Big Bang\vspace*{.5\baselineskip}}
\author{F.R. Klinkhamer}
\email{frans.klinkhamer@physik.uni-karlsruhe.de}
\affiliation{Institute for Theoretical Physics,\\
University of Karlsruhe (TH),\\76128 Karlsruhe, Germany} 

\begin{abstract}
\vspace*{.6\baselineskip}\noindent
The near-zero value of the cosmological constant $\Lambda$
in an equilibrium context may be due to the existence of a
self-tuning relativistic vacuum variable $q$.
Here, a cosmological nonequilibrium context is considered with a corresponding
time-dependent cosmological parameter $\Lambda(t)$
or vacuum energy density $\rho_\text{V}(t)$.
A specific model of a closed Friedmann--Robertson--Walker universe
is presented, which is determined by equilibrium boundary conditions
at one instant of time ($t=t_\text{eq}$)
and a particular form of vacuum-energy dynamics
($d\rho_\text{V}/dt \propto \rho_\text{M}$).
This homogeneous and isotropic model has a standard Big Bang phase at early
times ($t \ll t_\text{eq}$)
and reproduces the main characteristics of the present universe
($t=t_0 < t_\text{eq}$).
\end{abstract}

\maketitle

\section{Introduction}
\label{sec:Introduction}

It has been argued that the gravitating vacuum energy density
$\rho_\text{V}$ or cosmological constant
$\Lambda$~\cite{Einstein1917,Weinberg1972,HawkingEllis1973,Mukhanov2005,Weinberg2008}
vanishes in a perfect quantum vacuum, provided that
this vacuum can be considered to be a
self-sustained medium at zero external pressure and that
there exists a new type of conserved microscopic variable $q$ which
self-adjusts so as to give vanishing internal pressure~\cite{KlinkhamerVolovik2008-PRD77}.
As the perfect quantum vacuum is Lorentz invariant
(see, e.g., Refs.~\cite{KosteleckyMewes2002,KlinkhamerRisse2007}
for bounds on Lorentz violation in the photon sector),
this vacuum ``charge'' $q$ must be of an entirely new type,
different from known conserved
charges such as baryon number minus lepton number, $B-L$.
The detailed microscopic theory is, of course, unknown,
but two examples of possible theories with such a vacuum variable
$q$ have been given in Ref.~\cite{KlinkhamerVolovik2008-PRD77}.

For the perfect Lorentz-invariant quantum vacuum, the
vacuum variable $q$ is constant over the whole of
spacetime. The previous discussion then applies to an
equilibrium situation and describes what may be called
the ``statics of dark energy.'' Two outstanding
questions are, first, how the equilibrium argument relates to the observed
expanding universe and, second, which physical principle
governs the ``dynamics of dark energy.''
Obviously, these are profound questions
and the present article can only hope to provide a small step
towards a possible solution.
In fact, the first question is temporarily replaced by the following
restricted question:
\begin{quote}
Is it possible at all to relate equilibrium boundary
conditions for $\rho_\text{V}(t_\text{eq})$ to
an expanding universe which matches the observations,
even if we are free to choose the type of vacuum-energy dynamics,
$d\rho_\text{V}/dt\ne 0$?
\end{quote}
In mathematical terms, we are after an ``existence proof''
for this type of model universe, which has equilibrium boundary conditions
setting the numerical value of the vacuum energy density
$\rho_\text{V}$ at one moment in time
(here, coordinate time $t=t_\text{eq}\equiv 0$).

It turns out to be rather difficult to construct such
an existence proof, but, in the end, we have been able
to find one suitable class of universes. The main lesson
we will learn from this exercise is the  necessity of some form
of ``instability'' of the imperfect quantum vacuum (for the case
considered, Lorentz invariance is perturbed by the presence of thermal
matter and spatial curvature) and we will get an idea of what type of
instability would be required to reproduce the Universe as
observed~\cite{Weinberg1972,HawkingEllis1973,Mukhanov2005,Weinberg2008}.
In a way, our goal is to find the ``Kepler laws'' of the accelerating
universe, leaving the underlying physics to future generations.

The outline of this article is as follows.
The topic of dynamic vacuum energy density in the context of $q$--theory
is introduced in Sec.~\ref{sec:dynamic-vacuum-energy-from-q-theory}.
A closed Friedmann--Robertson--Walker (FRW)
universe with a generalized \emph{Ansatz} for the vacuum-energy
dynamics is then discussed in Sec.~\ref{sec:closed-FRW-universe-vacuum-dynamics}.
The corresponding numerical solution is presented
in Sec.~\ref{sec:nonstandard-universe}
(related results for the case of vanishing vacuum energy density
are relegated to the Appendix).
Final comments are given in Sec.~\ref{sec:Discussion}.

\section{Dynamic vacuum energy density from $\boldsymbol{q}$--theory}
\label{sec:dynamic-vacuum-energy-from-q-theory}

\subsection{Gravitational action with four-form and scalar fields}
\label{subsec:Gravitational-action-four-form-and-scalar-fields}

The crucial issue is the exchange of energy between the deep vacuum
(described in part by the conserved microscopic variable $q$)
and the low-energy degrees of freedom corresponding to the
physics of the standard model and general relativity.
The detailed microscopic theory is unknown, but we can try
to seek guidance from the concrete four-form theory
considered in Ref.~\cite{KlinkhamerVolovik2008-PRD77}.

This particular theory, coupled to low-energy matter, is defined by the
action~\cite{DuffNieuwenhuizen1980,Aurilia-etal1980,KlinkhamerVolovik2008-PRD77}
\begin{subequations}\label{eq:action+fourform}
\beqa
S&=& \int_{\mathbb{R}^4} \, d^4x \,\sqrt{-g}\,
\Bigg(\frac{R}{16\pi G_\text{N}}
- \frac{1}{2}\, \partial_\mu\phi\, \partial_\nu\phi\;g^{\mu\nu}
- \epsilon(F)\,\left(1+\frac{1}{2}\, \phi^2/M^2\right)
\Bigg)~,
\label{eq:action-matter}
\\[2mm]
F^2 &\equiv& -  \frac{1}{24}\,
F_{\mu\nu\rho\sigma}\, F_{\alpha\beta\gamma\delta}\:
g^{\alpha\mu}g^{\beta\nu}g^{\gamma\rho}g^{\delta\sigma}\,~,
\label{eq:fourformsquared}
\\[2mm]
F_{\mu\nu\rho\sigma} &\equiv& \nabla_{[\mu}A_{\nu\rho\sigma]}\,,
\label{eq:fourform-def}
\eeqa
\end{subequations}
where $R(x)$ is the Ricci curvature scalar from the metric $g_{\mu\nu}(x)$,
$F_{\mu\nu\rho\sigma}(x)$  the four-form field strength of
a three-form gauge field $A_{\nu\rho\sigma}(x)$, and
$\nabla_\mu$ the covariant derivative. In addition,
the microscopic energy density $\epsilon(F)$ is taken to be
an arbitrary function of $F$  and the low-energy matter field
$\phi(x)$ a real scalar field with coupling constant $1/M^2$ to
$\epsilon(F)$. Here, and in the rest of this section, we use natural units
with $\hbar=c=1$.

The variational principle applied to action \eqref{eq:action-matter}
results in three field equations,
a generalized Maxwell equation for the $F_{\mu\nu\rho\sigma}$ field,
a generalized Klein--Gordon equation for the $\phi$ field, and
the standard Einstein equation for the $g_{\mu\nu}$ field
with an energy-momentum tensor $T_{\mu\nu}$
from both $F_{\mu\nu\rho\sigma}$ and $\phi$ fields.

\subsection{Vacuum energy density in a flat FRW universe}
\label{subsec:vacuum-energy-flat-FRW-universe}

In order to solve the field equations from the model action
\eqref{eq:action+fourform}, the following \emph{Ansatz} can be used:
a spatially-flat ($k=0$) Friedmann--Robertson--Walker
metric~\cite{Weinberg1972}, a Levi--Civita-type four-form
field~\cite{DuffNieuwenhuizen1980,Aurilia-etal1980,KlinkhamerVolovik2008-PRD77},
and a homogenous scalar field. Specifically, the \emph{Ansatz} fields
are given by
\begin{subequations}\label{eq:gFPhi-Ansaetze}
\beqa
g_{\mu\nu}(x) &=& \diag\big(+1,-a(t)^2,-a(t)^2,-a(t)^2\big)~,
\label{eq:g-Ansatz}\\[1mm]
  F_{\mu\nu\rho\sigma}(x) &=& q(t)\;|a(t)|^3 \;e_{\mu\nu\rho\sigma}\,,
\label{eq:F-Ansatz}\\[1mm]
\phi(x)&=&\phi(t)~,
\label{eq:Phi-Ansatz}
\eeqa
\end{subequations}
with scale factor $a(t)$ and
totally antisymmetric Levi--Civita symbol $e_{\mu\nu\rho\sigma}$.
The generalized Maxwell equation reduces then to
\beq\label{eq:qdotoverq}
\frac{\dot{q}}{q}= -\chi_\text{V}\;\, q\,\epsilon^\prime\;\,
                   \frac{\phi \,\dot{\phi}}{M^2 +\phi^2/2}~,
\eeq
in terms of the vacuum compressibility~\cite{KlinkhamerVolovik2008-PRD77}
\beq\label{eq:chiV}
\chi_\text{V} \equiv \big(q^2\,\epsilon^{\,\prime\prime}\,\big)^{-1}\,,
\eeq
with the prime standing for differentiation with respect to the vacuum variable
$q$ and the overdot for differentiation with respect to
the cosmic time coordinate $t$.

With the \emph{Ansatz} fields \eqref{eq:gFPhi-Ansaetze},
there are two contributions to the energy-momentum tensor $T_{\mu\nu}$
in the Einstein field equation.
The first contribution to $T_{\mu\nu}$ is proportional
to the metric, $T_{\mu\nu}^\text{V}=\rho_\text{V}\,g_{\mu\nu}$,
and corresponds to a vacuum energy density
\begin{equation}
\rho_\text{V}= \Big(\epsilon(q) - q\, \epsilon^\prime(q)\Big)
\equiv
\widetilde{\epsilon}(q)\,,
\label{eq:Lambda-matter}
\end{equation}
which equals the previous result $\widetilde{\epsilon}(q)$
from Ref.~\cite{KlinkhamerVolovik2008-PRD77}.
In the following, it will be assumed that the equilibrium
value $q_\text{c}$ is such that $\widetilde{\epsilon}(q_\text{c})>0$
and $\chi_\text{V}(q_\text{c})>0$, where the value $q_\text{c}$
[different from the value $q_{0}$ for Minkowski spacetime with
$\widetilde{\epsilon}(q_0) = 0$]
may result from some type of perturbation
as discussed in Ref.~\cite{KlinkhamerVolovik2008-PRD77}.

The second contribution to $T_{\mu\nu}$ corresponds to
the energy-momentum tensor of a comoving perfect fluid with energy density and
pressure~\cite{HawkingEllis1973,Mukhanov2005,Weinberg2008}
given by
\beq\label{eq:epsilon-P}
\rho_\text{M} = \frac{1}{2}\, \dot{\phi}^2 + \frac{1}{2}\,\widetilde{\mu}^2\,\phi^2\,,
\quad
P_\text{M}    = \frac{1}{2}\, \dot{\phi}^2 - \frac{1}{2}\,\widetilde{\mu}^2\,\phi^2\,,
\eeq
in terms of the effective mass square
\beq\label{eq:widetilde-mu^2}
\widetilde{\mu}^2(q) \equiv \widetilde{\epsilon}(q)/M^2\,,
\eeq
which is positive as long as $\widetilde{\epsilon}(q)$ and $M^2$ are.
At the equilibrium value $q=q_\text{c}$, define
$\widetilde{\mu}_\text{c}^2 \equiv \widetilde{\mu}^2(q_\text{c})$.

For later, it turns out to be useful to introduce already the following
equations of state for matter and vacuum:
\beq\label{eq:EOS-MV}
P_\text{M} =w_\text{M}\, \rho_\text{M}~,\quad
P_\text{V} =w_\text{V}\, \rho_\text{V} =-\, \rho_\text{V}~.
\eeq
The matter equation-of-state parameter can be time dependent,
$w_\text{M}=w_\text{M}(t)$, as it is simply the ratio of the two terms in
\eqref{eq:epsilon-P}. But the vacuum equation-of-state parameter is
strictly constant, $w_\text{V}=-1$, as the corresponding energy-momentum
tensor~\cite{KlinkhamerVolovik2008-PRD77}
is given by $T_{\mu\nu}^\text{V}=\rho_\text{V}\,g_{\mu\nu}$.
This different behavior traces back to the special nature of
the four-form field without propagating degrees of
freedom~\cite{DuffNieuwenhuizen1980,Aurilia-etal1980}
and to the fact that there are no derivative
terms of $F$ in the original action \eqref{eq:action-matter}.

\subsection{Energy exchange between vacuum and matter}
\label{subsec:energy-exchange-between-vacuum-and-matter}

The structure of the vacuum energy density  \eqref{eq:Lambda-matter}
from the simple model considered allows us to say something
concrete about the energy exchange between vacuum and matter.
From the reduced Maxwell equation \eqref{eq:qdotoverq},
the time derivative of \eqref{eq:Lambda-matter} is
given by
\beq\label{eq:Lambdadot-Phidot}
\dot{\rho}_\text{V}
= \widetilde{\epsilon}^{\;\prime}\;\dot{q}
=-\widehat{\chi}\;\,
  \widetilde{\epsilon}\,\;\frac{\phi \,\dot{\phi}}{M^2 +\phi^2/2}~,
\eeq
for the dimensionless quantity
\beq\label{eq:widehat-chi}
\widehat{\chi} \equiv (q\, \widetilde{\epsilon}^{\;\prime}/\widetilde{\epsilon})
                       \;(q\,\epsilon^{\;\prime})
                       \;\big(q^2\;\epsilon^{\,\prime\prime}\big)^{-1}\,,
\eeq
whose absolute value may be of order 1 for generic $\epsilon(q)$.
Considering small field values $\phi^2 \ll M^2$ (see below),
the final expression reads
\beqa
\dot{\rho}_\text{V}(t)&=&
\text{sgn}\big[-\phi(t) \,\dot{\phi}(t)\big]\;\widehat{\mu}(t)\;
\big|\widetilde{\mu}_\text{c}\big|\;\sqrt{1-w_\text{M}^2(t)}\;\rho_\text{M}(t)\,,
\label{eq:Lambdadot-rhoM}
\eeqa
for a further dimensionless quantity
\beq\label{eq:widehat-mu}
\widehat{\mu}(t) \equiv \widehat{\chi}(t)\,
\big|\widetilde{\mu}\big(q(t)\big)\big| /\big|\widetilde{\mu}_\text{c}\big|\,,
\eeq
which can also be assumed to be of order 1, as long as $q(t)$ remains
close to $q_\text{c}$
[recall the definition of $\widetilde{\mu}_\text{c}^2$ a few lines
below \eqref{eq:widetilde-mu^2}].
For completeness, the sign function used in \eqref{eq:Lambdadot-rhoM} has
$\text{sgn}[x]\equiv x/|x|$ for $x\ne 0$ and $\text{sgn}[0]\equiv 0$.

At this moment, a brief comment on the energy scales involved may be
appropriate. The microscopic energy density $\epsilon(F)$ can be assumed to be
of order $(E_\text{Planck})^4$, with
$E_\text{Planck} \equiv \sqrt{\hbar\, c^5/G_\text{N}} \approx
1.22 \times 10^{28}\;\text{eV}$. In addition, if \eqref{eq:Lambdadot-rhoM}
is to play a role in the energy balance and evolution of the
present universe (see Sec.~\ref{sec:nonstandard-universe}),
one requires the following order of magnitudes~\cite{Weinberg2008}:
$\rho_\text{V} \sim \rho_\text{M} \sim (10^{-3}\;\text{eV})^4$ and
$\widetilde{\mu}_\text{c} \sim 10^{-33}\;\text{eV}$.
With the $\widetilde{\mu}_\text{c}^2$ definition
below \eqref{eq:widetilde-mu^2}, this gives
$M \sim \sqrt{\rho_\text{V}}/(10^{-33}\;\text{eV}) \sim 10^{27}\;\text{eV}$,
which corresponds to a Planckian energy scale. Of course, it remains to be seen
if such a toy-model version \eqref{eq:action-matter} of $q$--theory is relevant
to the ultimate microscopic theory.

To summarize, result \eqref{eq:Lambdadot-rhoM} describes the change of
vacuum energy density due to nontrivial matter dynamics ($\dot{\phi}\ne 0$)
and nonzero vacuum compressibility ($\chi_\text{V}>0$).
However, \eqref{eq:Lambdadot-rhoM} holds only
for matter described by a single real scalar field
$\phi$ and the flat ($k=0$) FRW universe.
More importantly, the dimensionless microscopic function $\widehat{\mu}(t)$
is not at all known, even if it can be expected to be of order unity.
In the following, we, therefore, work with an \emph{Ansatz} for
$\dot{\rho}_\text{V}$ which is kept as general as possible
but still proportional to $\rho_\text{M}$.

\section{Closed FRW universe and nontrivial vacuum dynamics}
\label{sec:closed-FRW-universe-vacuum-dynamics}

\subsection{Standard dynamics}
\label{subsec:standard-dynamics}

The spatially flat ($k=0$) FRW universe does not have
an obvious time for equilibrium boundary conditions,
apart from the limiting case with $a(t)\to \infty$ and $\rho_\text{M}(t)\to 0$
as $t\to \infty$. For this reason, we turn to the closed ($k=1$) FRW
universe~\cite{Weinberg1972} with metric
\beq
g_{00}(x) =1\,,\quad
g_{m0}(x) =0\,,\quad
g_{mn}(x) =-a^2(t)\;\widehat{g}_{mn}(x)\,,
\label{eq:g-ClosedFRW}
\eeq
in terms of the standard metric $\widehat{g}_{mn}$ of a unit 3--sphere
for spatial indices $m,n=1,2,3$.
The scale factor $a(t)$ now corresponds to the radius of the
closed universe and, as is well-known, can have a stationary point
at a finite value of $a(t)$.

Henceforth, we discuss only the dynamics of classical relativity
and use units with $c=8\pi G_\text{N}/3=1$, unless stated otherwise.
Note, however, that the boundary conditions to be presented
in Sec.~\ref{subsec:equilibrium-boundary-conditions} may rely
implicitly on quantum mechanics, as does the vacuum instability to be
discussed in Secs.~\ref{subsec:nonstatic-universe-from-vacuum-instability}
and \ref{subsec:additional-remarks}.

The dynamics of the standard closed ($k=1$) FRW universe~\cite{Weinberg1972} is
governed by the 00--component of the Einstein equation,
\beqa\label{eq:Einstein00Closed}
\ddot{a}/a &=&
-(4\pi G_\text{N}/3)\, \Big( \rho_\text{total}+3\, P_\text{total}\Big)
 =            
(8\pi G_\text{N}/3)\, \Big(\rho_\text{V}-\frac{1}{2}\, (1+3\, w_\text{M})\,\rho_\text{M}\Big)\,,
\eeqa
the energy-conservation equation,
\beq\label{eq:energy-conservationClosed}
\big( \dot{\rho}_\text{V}+\dot{\rho}_\text{M} \big) =
-3\,(\dot{a}/a)\,(1+w_\text{M})\,\rho_\text{M}\,,
\eeq
and a trivial vacuum-energy equation,
\beq\label{eq:tivial-vacuum-energy-density}
\dot{\rho}_\text{V}=0\,,
\eeq
which corresponds to the case of a genuine cosmological constant
(spacetime-independent vacuum energy density).
Equations \eqref{eq:Einstein00Closed} and \eqref{eq:energy-conservationClosed}
have been derived for equation-of-state (EOS) parameters
\beq\label{eq:EOS-parameters}
w_\text{M}=\text{const}\,,\quad
w_\text{V}=-1\,.
\eeq
Here, the matter EOS parameter has been assumed to
be time independent, but this assumption can be relaxed later.
The vacuum EOS parameter $w_\text{V}$ is to remain fixed to the
value $-1$, which is the case for $q$--theory~\cite{KlinkhamerVolovik2008-PRD77}
as mentioned a few lines below \eqref{eq:EOS-MV}.

Recall that, combined with energy conservation \eqref{eq:energy-conservationClosed},
the first-order Friedmann equation,
\beq\label{eq:FriedmannClosed}
\big(\dot{a}/a\big)^2 =
(8\pi G_\text{N}/3)\,\big( \rho_\text{V}+\rho_\text{M}\big)-k/a^2\,\big|_{k=1}\,,
\eeq
is equivalent~\cite{Weinberg1972} to
the second-order Einstein equation \eqref{eq:Einstein00Closed}, at least,
for appropriate boundary conditions.

\subsection{Static Einstein universe from equilibrium boundary conditions}
\label{subsec:equilibrium-boundary-conditions}

For the task outlined in Sec.~\ref{sec:Introduction}
(obtaining an ``existence proof''),
the \emph{static} Einstein universe~\cite{Einstein1917} suffices,
as it corresponds to an equilibrium state with constant radius $a$
and constant energy densities $\rho_\text{V}$ and $\rho_\text{M}$.
This static closed universe can simply be taken as the starting point of the
discussion in Sec.~\ref{subsec:nonstatic-universe-from-vacuum-instability},
but it is also possible to give an argument
for the two conditions that single out this particular universe from
other closed FRW universes.

In fact, the following two conditions can be seen to nullify the
right-hand sides of the differential
equations \eqref{eq:Einstein00Closed}, \eqref{eq:energy-conservationClosed},
and \eqref{eq:FriedmannClosed}.
The first condition makes sure that the expansion momentarily stops
($\dot{a}/a=0$) at the equilibrium point $t_\text{eq}\equiv 0$:
\beq\label{eq:FRWClosed-Friedmann-bc}
(8\pi G_\text{N}/3)\,\Big( \rho_\text{V}(t_\text{eq})
                  +\rho_\text{M}(t_\text{eq}) \Big)=
                  k\,a(t_\text{eq})^{-2}\,\Big|_{k=1}\,,
\eeq
with the gravitational coupling constant $G_\text{N}$ and
the dimensionless curvature parameter $k$ shown temporarily.
The second condition makes the acceleration or deceleration vanish
($\ddot{a}/a=0$) at the equilibrium point $t_\text{eq}\equiv 0$:
\beq\label{eq:FRWClosed-GibbsDuhem-bc}
\rho_\text{V}(t_\text{eq})=
w_\text{M}\,\rho_\text{M}(t_\text{eq}) +
\frac{1}{2}\,\big(1+w_\text{M}\big)\,\rho_\text{M}(t_\text{eq})\,,
\eeq
where, strictly speaking, $w_\text{M}$ stands for $w_\text{M}(t_\text{eq})$,
but, here, $w_\text{M}$ has been assumed constant.
Clearly, condition \eqref{eq:FRWClosed-Friedmann-bc} does not require a nonzero
value of the vacuum energy density, whereas \eqref{eq:FRWClosed-GibbsDuhem-bc} does,
provided the model universe contains matter. Historically,
this was indeed the reason for Einstein~\cite{Einstein1917}
to introduce his original cosmological constant, as he was aiming for
a static universe.

Let us briefly comment on a possible interpretation of this last condition,
which, in this context, was first discussed by Volovik~\cite{Volovik2005}.
The first term on the right-hand side of \eqref{eq:FRWClosed-GibbsDuhem-bc}
corresponds to the flat-spacetime
condition $\rho_\text{V}= P_\text{M}=w_\text{M}\,\rho_\text{M}$ from
pressure equilibrium $P_\text{V}+P_\text{M}=P_\text{ext}=0$
and the vacuum equation of state $P_\text{V} =-\, \rho_\text{V}$.
See Ref.~\cite{KlinkhamerVolovik2008-PRD77} for an extensive discussion of this
flat-spacetime result,
which traces back to a Gibbs--Duhem-type equation derived in $q$--theory.
The second term on the right-hand side of \eqref{eq:FRWClosed-GibbsDuhem-bc}
describes the gravitational effects, even though Newton's gravitational
constant $G_\text{N}$ does not appear explicitly in the final result
[note that $G_\text{N}$ does enter
condition \eqref{eq:FRWClosed-Friedmann-bc} explicitly].
Specifically, the complete relation \eqref{eq:FRWClosed-GibbsDuhem-bc} follows
from the two conditions $P_\text{V}+P_\text{M}+P_\text{grav}=0$
and $G_\text{N}\big(\rho_\text{V}+\rho_\text{M}+\rho_\text{grav}\big)=0$
for an effective gravitational equation of state
$P_\text{grav}=-(1/3)\,\rho_\text{grav}$.
See Sec.~7 of Ref.~\cite{Volovik2005} for further discussion of this
curvature contribution to \eqref{eq:FRWClosed-GibbsDuhem-bc}.

Conditions \eqref{eq:FRWClosed-Friedmann-bc} and
\eqref{eq:FRWClosed-GibbsDuhem-bc} are simply boundary conditions for the
cosmological equations \eqref{eq:Einstein00Closed}--\eqref{eq:FriedmannClosed}.
But this last condition \eqref{eq:FRWClosed-GibbsDuhem-bc}
can also be argued from thermodynamic principles~\cite{Volovik2005}
and, for the quantum vacuum as discussed in Ref.~\cite{KlinkhamerVolovik2008-PRD77}
and Sec.~\ref{sec:dynamic-vacuum-energy-from-q-theory} here,
would have a naturally small vacuum energy density by the self-adjustment
of the vacuum variable $q$. The self-adjustment may, in fact,  be the result
from a very long phase in the ``life'' of the model universe, as will be
discussed in Sec.~\ref{subsec:additional-remarks}.

\subsection{Nonstatic universe from vacuum instability}
\label{subsec:nonstatic-universe-from-vacuum-instability}

Given the boundary conditions \eqref{eq:FRWClosed-Friedmann-bc} and
\eqref{eq:FRWClosed-GibbsDuhem-bc} and given the task
of somehow recovering the observed (expanding!) Universe,
the problem is to get \emph{away} from the static
Einstein universe~\cite{Einstein1917}
with $a(t)=a(0)$, $\rho_\text{V}(t)=\rho_\text{V}(0)$, and
$\rho_\text{M}(t)=\rho_\text{M}(0)$.
It appears that the only way to achieve this is
to consider either a modification of gravity
(e.g., a modified Einstein field equation as studied in Ref.~\cite{Volovik2003}) or
a new type of instability of the imperfect quantum vacuum.
The present article follows the second approach.

Specifically, we assume that \eqref{eq:tivial-vacuum-energy-density}
is replaced by the following \emph{Ansatz} for
the time variation of the vacuum energy density:
\beq
\dot{\rho}_\text{V}(t) = \GammaVM\,\gammaV(t)\;\rho_\text{M}(t)\,,
\label{eq:rhoVdot-Ansatz}
\eeq
with a dimensionless functional $\gammaV[a(t)/a_\text{eq}]\equiv \gammaV(t)$,
normalized by $\gammaV[1]=1$, and a new fundamental decay constant
$\GammaVM>0$ [here, quantum mechanics may enter if,
for example, $\GammaVM \propto m c^2/\hbar$ for a mass scale $m$,
as in \eqref{eq:Lambdadot-rhoM} from the simple version of $q$--theory
discussed in Sec.~\ref{sec:dynamic-vacuum-energy-from-q-theory}].
As mentioned before, the origin of \eqref{eq:rhoVdot-Ansatz}
needs to be explained by the detailed microphysics, but, here,
we take a purely phenomenological (``Keplerian'') approach
and simply assume a particular form for $\dot{\rho}_\text{V}$.
Remark that, for $w_\text{M}=0$, $\rho_\text{M}$ in \eqref{eq:rhoVdot-Ansatz}
can be interpreted as corresponding to the cold-dark-matter energy density
from observational cosmology~\cite{Weinberg2008},
with the baryonic contribution neglected.

Equations \eqref{eq:Einstein00Closed}, \eqref{eq:energy-conservationClosed},
and \eqref{eq:rhoVdot-Ansatz}
with boundary conditions \eqref{eq:FRWClosed-Friedmann-bc} and
\eqref{eq:FRWClosed-GibbsDuhem-bc} can then be solved numerically
to give $a(t)$, $\rho_\text{M}(t)$, and  $\rho_\text{V}(t)$.
As we intend to take equilibrium-point boundary conditions also for the
standard case with $\rho_\text{V}(t)=0$ [some relevant results
are given in the Appendix],
we use the second-order 00--component Einstein equation \eqref{eq:Einstein00Closed}
instead of the first-order Friedmann equation \eqref{eq:FriedmannClosed}.
As mentioned before, it is a well-known fact~\cite{Weinberg1972} that,
with appropriate boundary conditions,
the differential equations \eqref{eq:Einstein00Closed} and
\eqref{eq:FriedmannClosed} are equivalent
when combined with the energy-conservation equation
\eqref{eq:energy-conservationClosed}. Incidentally,
the 11--component of the Einstein equation is also satisfied,
as are the 22 and 33 components by isotropy.

\subsection{Additional remarks}
\label{subsec:additional-remarks}

In this subsection, further remarks are presented on the
background and context of the vacuum-instability \emph{Ansatz}
\eqref{eq:rhoVdot-Ansatz}. These remarks are, however, not essential
for the rest of this article and can be skipped in a first reading.

To start, three technical remarks on \emph{Ansatz} \eqref{eq:rhoVdot-Ansatz}
are in order:
\begin{enumerate}
\item
$\dot{\rho}_\text{V}$ vanishes if $\rho_\text{M}=0$, but $\rho_\text{V}$
can still be nonzero, so that a de-Sitter universe remains a possible solution
for the case of $\rho_\text{M}=0$ and $\rho_\text{V}=\text{const}>0\,$;
\item
$\dot{\rho}_\text{V}$ does not necessarily vanish if $\dot{a}/a=0$ and,
in particular, $\dot{\rho}_\text{V}$ does not vanish at $t=t_\text{eq}\equiv 0$,
so that the model universe can get away from the static Einstein universe;
\item
time-reversal invariance around $t_\text{eq}$ is manifestly broken
if $\gammaV(t)$ is continuous at $t=t_\text{eq}$.
\end{enumerate}
Note that \emph{Ansatz} \eqref{eq:rhoVdot-Ansatz}
resembles Eq.~(8) of Ref.~\cite{Barcelo2007}
with $\Gamma_\text{VV} \equiv 1/\tau  \ne 0$
and Eq.~(3) of Ref.~\cite{Amendola2007},
but differs by points 1 and 2, respectively.
Observe also that point 1 holds precisely for result
\eqref{eq:Lambdadot-rhoM} derived from the toy-model version of $q$--theory
in Sec.~\ref{sec:dynamic-vacuum-energy-from-q-theory}.

From point 2 above and with $\GammaVM\, \gammaV(t_\text{eq})>0$,
there is, in principle, the possibility of having a ``Big Bang''
with $a(t_\text{BB})=0$ at $t_\text{BB} < t_\text{eq}$.
Remark that the direction of the coordinate time $t$ has no direct physical meaning
for the homogenous models considered here, as
the physical ``arrow-of-time'' appears to be related to
the ``growth'' of inhomogeneities ``originating'' from a
smooth Big Bang~\cite{Penrose1979}
(see also Ref.~\cite{Klinkhamer2002} for an explicit T--violation mechanism
in a closed nonisotropic universe).

From point 3, there is  the possibility that,
even with a Big Bang at $t_\text{BB} < t_\text{eq}$,
the model universe does \emph{not} return to vanishing 3--volume for $t> t_\text{eq}$.
One possible scenario is that the function $\gammaV(t)$ has
a discontinuous jump to $\gammaV(t)=0$
for $t> t_\text{eq}$ and that the homogeneous model universe is static for
$t \in [t_\text{eq},\infty)$.
There would then be an infinitely long equilibrium phase which makes the
discussion of an self-adjusting vacuum variable $q$ quite
natural~\cite{KlinkhamerVolovik2008-PRD77}
(the vacuum variable $q$ may also play a crucial role for the stability
issue; see Sec.~II C of Ref.~\cite{KlinkhamerVolovik2008-PRD77}).
Considering the coordinate time $t$ to ``start'' at a large
positive value and to ``run'' in the negative direction,
the nonstatic universe then ``takes off'' at
$t\equiv 0$ due to the sudden onset of instability, leading to a ``Big Bang''
for an appropriate behavior of $\gammaV(t)$ at $t \leq 0$,
as will be discussed in the next section.
This fluctuation scenario resembles, in a way,
earlier discussions~\cite{Vilenkin1984-1999} on the tunneling origin
of the nonstatic universe (around $a \sim 0$),
but our fluctuation ``starts at the other end,'' that is,
$a \sim a_\text{eq}$.

As a final remark, we emphasize that the model considered in the present article
is based on the Gibbs--Duhem-type condition \eqref{eq:FRWClosed-GibbsDuhem-bc}
of the static Einstein universe, which may arise from the self-adjustment of a
conserved relativistic variable $q$ characterizing the
microscopic quantum vacuum~\cite{KlinkhamerVolovik2008-PRD77}.
In this respect, the closed model universe presented here
is complementary to the model of a scalar field evolving towards an attractor
(see, e.g., Ref.~\cite{RatraPeebles1988} and references therein), as this
type of scalar model does not solve
the quantum-mechanical cosmological constant problem
of why $\rho_\text{V}$ vanishes in Minkowski spacetime without fine tuning.
The general analysis of an evolving scalar field may, indeed, turn out to be
relevant for an accurate description of the present universe
with $\rho_\text{V} \sim \rho_\text{M} \ll E_\text{Planck}^4$,
especially if the effective scalar field can be related to a
conserved microscopic variable $q$.
In the present article, however, we do not considered the dynamics of
$q$ or other microscopic fields and use, instead,
the simple phenomenological  \emph{Ansatz} \eqref{eq:rhoVdot-Ansatz}.

\section{Nonstandard closed FRW universe}
\label{sec:nonstandard-universe}

\subsection{Specific $\boldsymbol{\gammaV}$ Ansatz}
\label{subsec:specific-gamma-Ansatz}

As explained in the Introduction, our goal is relatively modest:
to find at least one functional $\gammaV[a(t)/a_\text{eq}]$ so that
Eqs.~\eqref{eq:Einstein00Closed}, \eqref{eq:energy-conservationClosed}
and \eqref{eq:rhoVdot-Ansatz},
with boundary conditions \eqref{eq:FRWClosed-Friedmann-bc} and
\eqref{eq:FRWClosed-GibbsDuhem-bc}
can produce a solution which more or less reproduces our known Universe
(see, e.g., Refs.~\cite{Weinberg2008,Eisenstein2005,Astier2006,Riess2007,Komatsu2008}
and references therein),
which is spatially flat to a high degree of precision and
approximately consists of $75\,\%$ ``dark energy'' and
$25\,\%$ matter (primarily nonbaryonic ``cold dark matter'').

With three coupled nonlinear ordinary differential equations (ODEs),
this modest goal is surprisingly difficult to reach.
Still, we have been successful by first considering the \emph{inverse}
problem which consists of the following two steps:
(i) to find, given a more or less reasonable $a_\text{designer}(t)$,
which densities $\rho_\text{M}(t)$ and $\rho_\text{V}(t)$ are required;
(ii) to determine, by differentiation of the $\rho_\text{V}(t)$ from
the first step, the required $\GammaVM\,\gammaV(t)$ from \eqref{eq:rhoVdot-Ansatz}.

Inspired by these ``designer-universe'' results, we make the following
\emph{Ansatz} for the (dimensionless) vacuum-dynamics functional:
\begin{subequations}\label{eq:gammaAnsatz}
\beqa\hspace*{-1cm}
\gammaV\left[\alpha(t)\right]&=&
        \alpha^2 \,f_{c_1}(1-\alpha)\; \sin\left(c_2\, \pi\, \alpha\right)
+ \alpha \,f_{c_1}(\alpha)\,
  \left(\frac{(c_3)^{1/3}}{(c_3)^{1/3} +|1-\alpha|^{1/3}}\right)^4\,,
\\[2mm]
f_{c}(x) &\equiv& x^6\,\left(1+c^6\right)/\left(x^6+c^6\right)\,,
\eeqa\end{subequations}
with $\alpha(t)\equiv a(t)/a_\text{eq}$ restricted to the range $[0,1]$
and numerical coefficients $c_n > 0$.
Roughly speaking, this \emph{Ansatz} for $\gammaV(t)$
consists of a sharply-peaked positive term
modulated to be effective just below $a=a_\text{eq}$ and
a term proportional to $a^3$ modulated to be effective near $a=0$.
A nonzero value of $\gammaV(t_\text{eq})$ will be seen to be needed
to get a nonstatic universe and the behavior $\gammaV\propto a^3$ near
$a=0$ will be seen to allow for a finite limiting value of $\rho_\text{V}(a)$
by compensating the divergent $w_\text{M}=0$ behavior
$\rho_\text{M}\propto 1/a^3$
on the right-hand side of \eqref{eq:rhoVdot-Ansatz}.

\subsection{Numerical solution}
\label{subsec:numerical-solution}

%
\newcommand{\cONEnum}        {1/5}
\newcommand{\cTWOnum}        {9/4}
\newcommand{\cTHREEnum}      {1/15}
\newcommand{\tBBnum}         {-0.91636}
\newcommand{\ABStBBnum}      {0.91636}
\newcommand{\tNOWnum}        {-0.5842}
\newcommand{\aNOWnum}        {5.582\phantom{0}}
\newcommand{\HubbleNOWnum}   {2.985\phantom{0}}
\newcommand{\rhoMNOWnum}     {2.384\phantom{0}}
\newcommand{\rhoVNOWnum}     {6.557\phantom{0}}
\newcommand{\rhoVoverMNOWnum}{2.750\phantom{0}}
\newcommand{\OmegaVMNOWnum}  {1.004\phantom{0}}
\newcommand{\tNOWminustBBnum}{0.3322}    
\newcommand{\tauNOWnum}{13.85}           
\newcommand{\taueqnum} {38.22}           
\newcommand{\aRELnum}  {6\times 10^{-3}} 
\newcommand{\tzONE}    {-0.75}           
%

A concrete model universe can be obtained by taking the following numerical
values (in units with $8\pi G_\text{N}/3=c=1$) for the boundary conditions
at $t=t_\text{eq}\equiv 0$ and the model parameters
(namely, the matter EOS parameter $w_\text{M}$,
the vacuum decay constant $\GammaVM$, and the \emph{Ansatz} coefficients $c_n$):
\beq\label{eq:BCSnonstandarduniverse}
\left(
  \begin{array}{c}
    a(0) \\
    \rho_\text{M}(0) \\
    \rho_\text{V}(0) \\
    w_\text{M}\\
    \GammaVM \\
    c_1 \\
    c_2 \\
        c_3 \\
  \end{array}
\right)
=
\left(
  \begin{array}{c}
    10\\
    2/300\\
    1/300 \\
    0\\
    50\\
    \cONEnum\\
    \cTWOnum\\
    \cTHREEnum\\
    \end{array}
\right)\,,
\eeq
with the implicit equilibrium condition
$\dot{a}/a=0$ at $t=0$ from the Friedmann equation \eqref{eq:FriedmannClosed}.
Remark that, if, for example, the value of the curvature radius $a(0)$
is fixed, the values of the energy densities
$\rho_\text{M}(0)$ and $\rho_\text{V}(0)$ are determined by
the equilibrium conditions \eqref{eq:FRWClosed-Friedmann-bc}
and \eqref{eq:FRWClosed-GibbsDuhem-bc}, for given $w_\text{M}$.

\begin{figure*}[t]  
\includegraphics[width=1\textwidth]{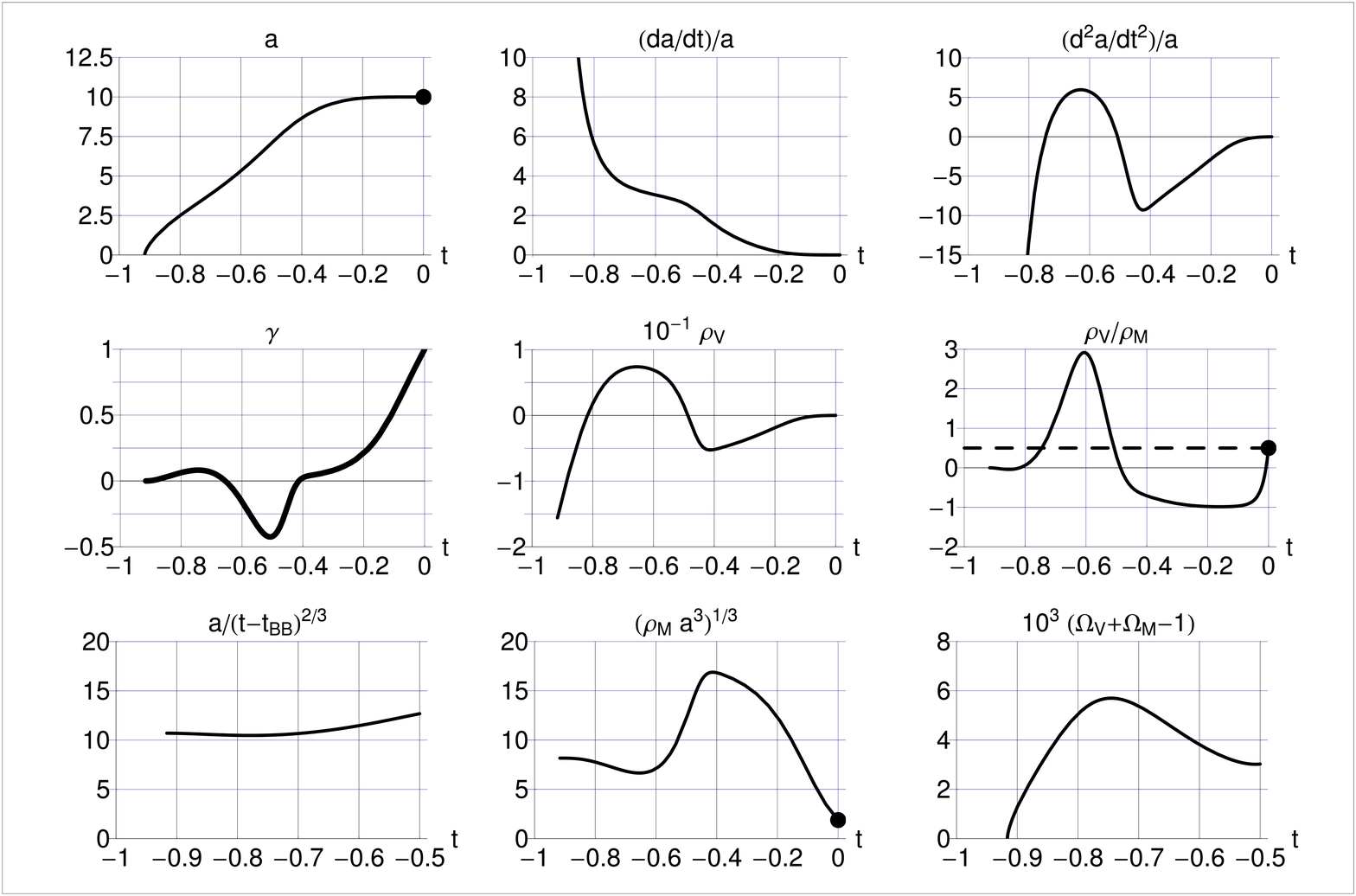} 
\caption{Closed FRW universe with pressureless matter
($w_\text{M}=0$) and dynamic vacuum energy ($w_\text{V}=-1$),
for boundary conditions \eqref{eq:BCSnonstandarduniverse}
in units with $8\pi G_\text{N}/3=c=1$.
The assumed behavior of the vacuum-energy dynamics is given by
\eqref{eq:rhoVdot-Ansatz} with the functional $\gammaV[a(t)/a_\text{eq}]$
from \eqref{eq:gammaAnsatz}.
The three nonzero equilibrium boundary conditions
on $a$, $\rho_\text{M}$, and $\rho_\text{V}$ at $t= t_\text{eq}\equiv 0$
are indicated by the heavy dots (only shown if clearly different from zero)
and the functional $\gammaV$ (with particular values for the numerical
coefficients $c_1$, $c_2$, and $c_3$)
is indicated by the heavy curve in the left-most panel of the middle row.
Moreover, the vacuum decay constant $\GammaVM$ has been set to $50$
and this relatively large value explains the rapid change
of $\rho_\text{V}/\rho_\text{M}$ near $t=0$.
The scale factor $a(t)$ vanishes at $t=t_\text{BB}=\tBBnum$
and the expansion of the model universe is accelerated ($\ddot{a}/a>0$) if
$\rho_\text{V}/\rho_\text{M} > 1/2$, as indicated by the dashed
curve in the right-most panel of the middle row.}
\label{fig:ClosedFRWgammaAnsatz}
\end{figure*}

The numerical solution of the coupled ODEs
\eqref{eq:Einstein00Closed}, \eqref{eq:energy-conservationClosed},
and \eqref{eq:rhoVdot-Ansatz} with \emph{Ansatz} \eqref{eq:gammaAnsatz}
and boundary conditions \eqref{eq:BCSnonstandarduniverse}
is given in Fig.~\ref{fig:ClosedFRWgammaAnsatz}.
[It has been verified explicitly that this numerical solution also
solves the Friedmann equation \eqref{eq:FriedmannClosed}.]
Observe that $\GammaVM=0$ would give a static Einstein universe with
$a(t)=a(0)$, $\rho_\text{V}(t)=\rho_\text{V}(0)$, and
$\rho_\text{M}(t)=\rho_\text{M}(0)$ at the values indicated by
the heavy dots in Fig.~\ref{fig:ClosedFRWgammaAnsatz}.
As explained in Sec.~\ref{subsec:nonstatic-universe-from-vacuum-instability}
and \ref{subsec:additional-remarks},
we have appealed to a new type of ``instability'' of the imperfect quantum
vacuum with $\GammaVM\,\gammaV(t_\text{eq}) > 0$ in order
to get away from this static universe
(for time coordinate $t$ starting at a value 0 and running in the
negative direction, so that $\rho_\text{V}$ decreases initially).

\subsection{Big Bang and present universe recovered}
\label{subsec:Big-Bang-and-present-universe-recovered}

Turning to the detailed model results of Fig.~\ref{fig:ClosedFRWgammaAnsatz},
the ``Big Bang'' with $a(t_\text{BB})=0$
would occur at coordinate time $t=t_\text{BB}=\tBBnum$, which differs
by 1 order of magnitude from the result without vacuum energy in the Appendix.
Still, approximately the same behavior for $t \downarrow t_\text{BB}$
is observed for both model universes, namely,
a scale factor vanishing as $a(t) \propto (t-t_\text{BB})^{2/3}$ and a matter
energy density diverging as $\rho_\text{M} \propto a^{-3}$, with the vacuum
energy density $\rho_\text{V}(t)$ in Fig.~\ref{fig:ClosedFRWgammaAnsatz}
approaching a finite value at $t= t_\text{BB}$
[cf. the last sentence of Sec.~\ref{subsec:specific-gamma-Ansatz}].

The ``present universe'' with density ratio
$\rho_\text{V}/\rho_\text{M} \approx 2.75$
(close to the WMAP--5yr mean value from Table 1 in Ref.~\cite{Komatsu2008}
for $h=0.70$) would approximately correspond to the time $t=t_0=\tNOWnum$
in Fig.~\ref{fig:ClosedFRWgammaAnsatz}
(choosing the latest time of two possible times, which both happen to be
close to the maximum of $\rho_\text{V}/\rho_\text{M}$).
The model values of the present universe are then
\beq\label{eq:presentuniverse}
\left(
  \begin{array}{c}
    t\\
    t-t_\text{BB}\\
    a\\
    \dot{a}/a\\
    \rho_\text{M}\\
    \rho_\text{V}\\
    \rho_\text{V}/\rho_\text{M}\\
    \Omega_\text{V}+\Omega_\text{M}\\
  \end{array}
\right)
=
\left(
  \begin{array}{r}
    \tNOWnum\\
    \tNOWminustBBnum\\
    \aNOWnum\\
    \HubbleNOWnum\\
    \rhoMNOWnum\\
    \rhoVNOWnum\\
    \rhoVoverMNOWnum\\
    \OmegaVMNOWnum\\
  \end{array}
\right)\,,
\eeq
where $\Omega_\text{X}$ is the energy density $\rho_\text{X}$
relative to the critical density $\rho_\text{crit}\equiv (\dot{a}/a)^2$
in units with $8\pi G_\text{N}/3=c=1$

By identifying the calculated value $\dot{a}/a=\HubbleNOWnum$
with the measured value~\cite{Freedman2001} of the Hubble constant
$H_0 \equiv h/(9.78\times 10^{9}\;\text{yr}) \approx 0.70/(9.78\;\text{Gyr})$,
the present age of the model universe
$t_0-t_\text{BB}\approx \tNOWminustBBnum$ becomes
\beq\label{eq:tau0}
\tau_0\approx \tauNOWnum\;(0.70/h)\;\text{Gyr}\,.
\eeq
Similarly, the present radius of the model  universe $a_0\approx \aNOWnum$
becomes of the order of $2 \times 10^{11}\;\text{lyr}$,
significantly larger than the present particle horizon.
It is far from trivial that more or less reasonable values for
$\rho_\text{V0}/\rho_\text{M0}$, $\Omega_{\text{V}0}+\Omega_{\text{M}0}$,
and $\tau_0$ can be produced at all in our approach.

The equilibrium time $t_\text{eq}-t_\text{BB} \approx \ABStBBnum$ of the
model universe corresponds to $\tau_\text{eq} \approx  \taueqnum\;\text{Gyr}$,
but there need not be a Big Crunch at even later times
because of the possible lack of time-reversal invariance.
In fact, there may be a very long static phase with $a(t)=a(0)$
for $t\geq 0$, if $\gamma(t)=0$ for positive times $t$.
This possibility has already been discussed in the
penultimate paragraph of Sec.~\ref{subsec:additional-remarks}.

With the measured photon temperature
$T_{\gamma 0}\approx 3\,\text{K}$ and the model value $a_0 \approx 6$,
the matter EOS parameter $w_\text{M}$ must change to a value 1/3
for $0 \leq  a \lesssim\aRELnum$
(relativistic matter being dominant for $T \gtrsim 3000\,\text{K}$),
in order to recover the standard nucleosynthesis of the very early
universe~\cite{Weinberg1972,Mukhanov2005}.
In order to maintain a finite $\rho_\text{v}$ value as $a \downarrow 0$,
the \emph{Ansatz} \eqref{eq:gammaAnsatz} can have the factor
$\alpha^2$ in the first term on the right-hand side changed to $\alpha^3$,
for example, so that $\gamma \propto a^4$ for very small values of $a$.

As to the phenomenology of $\gammaV[a(t)/a_\text{eq}]$, we clearly
recognize three phases in Fig.~\ref{fig:ClosedFRWgammaAnsatz},
where $\gammaV(t)$ is positive, negative, and again positive
as the time coordinate $t$ moves away from $t_\text{eq}=0$
in the negative direction.
(Other structures of $\gammaV$ are not excluded \emph{a priori},
but the one found suffices for the present discussion.)
The resulting behavior of
$\rho_\text{V}(t)$ from \eqref{eq:rhoVdot-Ansatz}
is shown in the figure panel to the right of the one of $\gammaV(t)$.
The fact that there is energy exchange between vacuum and matter
is demonstrated by the nonconstant behavior of $\rho_\text{M}\,a^3$
as shown by the middle panel of the bottom row
of Fig.~\ref{fig:ClosedFRWgammaAnsatz}
(compare with the results in the Appendix).
Note that $\rho_\text{V}(t_\text{BB})$ need not be negative, as
different $\GammaVM$ values and coefficients $c_n$
in \eqref{eq:gammaAnsatz} can give
positive $\rho_\text{V}(t_\text{BB})$  values of order 1 or
perhaps $\rho_\text{V}(t_\text{BB})=0$.
Different $\GammaVM$ values and coefficients $c_n$ can also give a
$\rho_\text{V}/\rho_\text{M}$ peak value larger than 3,
but it may be difficult to keep the ``present age''
of the model universe at the value \eqref{eq:tau0} and to prevent it
from dropping to a significantly lower value.

From the approximate linearity of $a(t)$
up to the ``present value'' $t_0\approx \tNOWnum$ in
Fig.~\ref{fig:ClosedFRWgammaAnsatz},
it is possible to relate
the time coordinate $t$ just below $t_0$
to the redshift $z$ used by observational cosmology through
the approximate relation $1+z \approx (t_0-t_\text{BB})/(t-t_\text{BB})$.
Then, a coordinate time $t=\tzONE$
would correspond to a redshift $z \approx 1$
and the model vacuum energy density $\rho_\text{V}(z)$
from Fig.~\ref{fig:ClosedFRWgammaAnsatz} is seen to be
more or less constant for redshifts $z$ between $0$ and $1$.
In fact, if future observations can measure
$\rho_\text{V}(z)$ and $\rho_\text{M}(z)$ up to $z\approx 3$
(see, e.g., Ref.~\cite{SahniStarobinsky2006} for theoretical considerations),
this would indirectly constrain the behavior of $\GammaVM\,\gammaV(t)$ for
$t \in  (t_\text{BB},\, t_0]$.
These observations can perhaps also constrain $\GammaVM\,\gammaV(t)$
over the \emph{whole} range $[t_\text{BB},\,t_\text{eq}]$ if there are
effective two-boundary conditions such as $\rho_\text{V}(t_\text{BB})=0$
and $\rho_\text{V}(t_\text{eq})\ne 0$ from the underlying
microscopic physics (possibly with a new mechanism of
T and CPT violation~\cite{Penrose1979,Klinkhamer2002}).

\section{Discussion}
\label{sec:Discussion}

By way of summary, we list the main features of the particular
closed model universe of Sec.~\ref{sec:nonstandard-universe} and
Fig.~\ref{fig:ClosedFRWgammaAnsatz}:
\begin{enumerate}
\item
a Gibbs--Duhem-type boundary condition \eqref{eq:FRWClosed-GibbsDuhem-bc}
at $t=t_\text{eq}$ with a finite vacuum energy density
$\rho_\text{V}(t_\text{eq})=(1/2) \; \rho_\text{M}(t_\text{eq})$
for matter with equation-of-state parameter $w_\text{M}=0$
[this particular value for $\rho_\text{V}(t_\text{eq})$ may result from
the self-tuning of a conserved microscopic variable $q$
to an equilibrium value $q_\text{c}$];
\item
finite $|\rho_\text{V}(t)|$ within a factor of order
$10^3$ from the value set at $t=t_\text{eq}$ (see point 1);
\item
a standard Big Bang phase at $t \sim t_\text{BB} < t_\text{eq}$
having $a(t) \propto (t-t_\text{BB})^{2/3}$
for $w_\text{M}=0$, matter energy density $\rho_\text{M} \propto a^{-3}$,
and energy density ratio
$\rho_\text{V}/\rho_\text{M}\to 0$ for $t \downarrow t_\text{BB}$;
\item
an accelerating phase for ``present times,''
with $\rho_\text{V}/\rho_\text{M}$ of order 1
and an approximately flat 3--geometry.
\end{enumerate}
This  model universe constitutes the ``existence proof'' announced
in Sec.~\ref{sec:Introduction}.
Points 1 and 2 suggest, moreover, that a nonvanishing vacuum energy
density $\rho_\text{V}(t)$ relevant to cosmology
may not require fine-tuning by factors of
order $(E_\text{Planck}/10^{-3}\;\text{eV})^4 \sim 10^{124}$
due to the self-adjustment~\cite{KlinkhamerVolovik2008-PRD77}
of the vacuum variable $q$ in an equilibrium phase $t \geq t_\text{eq}$.

Still, it remains to be explained theoretically
that the fundamental vacuum-dynamics constant
$c/\GammaVM \approx 1 \times 10^{9}\;\text{lyr}
\approx \hbar c/(2 \times 10^{-32}\:\text{eV})$ is
of the order of the length scale
$a(t_\text{eq})\approx 4 \times 10^{11}\;\text{lyr}$
of the equilibrium model universe.
[As mentioned before, the single quantity $a(t_\text{eq})$
determines the two other quantities
$\rho_\text{M}(t_\text{eq})$ and $\rho_\text{V}(t_\text{eq})$ from
conditions \eqref{eq:FRWClosed-Friedmann-bc} and \eqref{eq:FRWClosed-GibbsDuhem-bc}
for given value of $w_\text{M}$.]
The theoretical explanation of this very small energy scale
$\hbar\,\GammaVM \approx 2 \times 10^{-32}\:\text{eV}$
would, most likely, trace back to the detailed microphysics, perhaps
along the lines of the simple version of $q$--theory
discussed in Sec.~\ref{subsec:energy-exchange-between-vacuum-and-matter}.
Inversely, there is the possibility that
observational cosmology, by measuring the time dependence of
the vacuum energy density, can provide
information on the microscopic structure of the quantum vacuum.

\vspace*{-1mm}
\section*{ACKNOWLEDGMENTS}

It is a pleasure to thank G.E. Volovik and J. Weller for valuable discussions
and the referee for helpful comments.

\vspace*{-1mm}
\section*{NOTE ADDED}

The $q$--theory approach~\cite{KlinkhamerVolovik2008-PRD77}
to dynamical vacuum energy density in cosmology
has been elaborated in two recent
articles~\cite{KlinkhamerVolovik2008-PRD78,KlinkhamerVolovik2008-JETPL}.

\vspace*{-1mm}
\begin{appendix}
\section{Standard closed FRW universe}
\label{app:standard-closed-FRW-universe}

In this appendix, a standard closed FRW universe~\cite{Weinberg1972}
is reviewed which has the same extremal radius as the nonstandard
universe discussed in Sec.~\ref{sec:nonstandard-universe}.
Specifically, the boundary conditions at $t=t_\text{max}\equiv 0$
and the matter equation-of-state parameter $w_\text{M}$
are (in units with $8\pi G_\text{N}/3=c=1$):
\beq\label{eq:BCSstandarduniverse}
\left(
  \begin{array}{c}
    a(0) \\
    \rho_\text{M}(0) \\
    \rho_\text{V}(0) \\
    w_\text{M}\\
      \end{array}
\right)
=
\left(
  \begin{array}{c}
    10\\
    1/100\\
    0 \\
    0\\
   \end{array}
\right)\,.
\eeq
Note that boundary conditions \eqref{eq:BCSstandarduniverse}
imply $\dot{a}/a=0$ at $t=0$ by the Friedmann equation \eqref{eq:FriedmannClosed}.

\begin{figure*}[b]  
\includegraphics[width=1\textwidth]{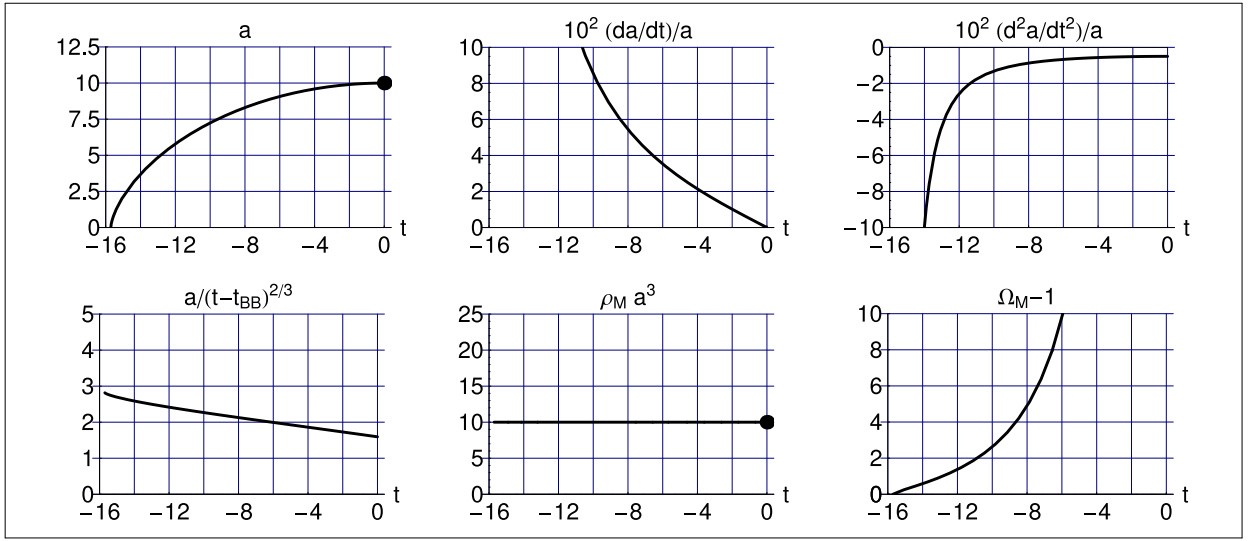}
\caption{Closed FRW universe with pressureless-matter energy density
$\rho_\text{M}(t)$ and vanishing vacuum energy density $\rho_\text{V}(t)$
[not displayed], for boundary conditions \eqref{eq:BCSstandarduniverse} in
units with $8\pi G_\text{N}/3=c=1$. On the first row are shown
the scale factor $a(t)$ and various derivatives,
$\dot{a}/a$ and $\ddot{a}/a$.
On the second row are shown
$a(t)$ scaled by a fractional power of the elapsed time since
$t_\text{BB}=-5\,\pi \approx -15.71$ where
$a(t)$ vanishes, the matter energy density $\rho_\text{M}$
multiplied by $a^3$, and the matter-density parameter
$\Omega_\text{M}\equiv \rho_\text{M}/\rho_\text{crit}$ defined in
terms of the critical density $\rho_\text{crit}\equiv (\dot{a}/a)^2$.
The two boundary conditions on $a$ and $\rho_\text{M}$
at $t=t_\text{max}\equiv 0$ are indicated by heavy dots.\vspace*{0cm}}
\label{fig:ClosedFRWstandard}
\end{figure*}

The corresponding numerical solution of the differential
equations \eqref{eq:Einstein00Closed}, \eqref{eq:energy-conservationClosed},
and \eqref{eq:tivial-vacuum-energy-density}
is displayed in Fig.~\ref{fig:ClosedFRWstandard}.
The analytic solution, in terms of an auxiliary angle $\theta \in [0,2\pi]$,
is given by~\cite{Weinberg1972,HawkingEllis1973}
\begin{subequations}\label{eq:StandardClosedUniverse}
\beqa
a&=&a_\text{max}\,\sin^2(\theta/2)\,,
\label{eq:StandardClosedUniverse-a}\\[2mm]
\rho_\text{M}(a)&=&a_\text{max}/a^3\,,\quad
\rho_\text{V}(a)=0\,,
\label{eq:StandardClosedUniverse-rhoMV}\\[2mm]
t&=&(\theta-\sin\theta-\pi)\,a_\text{max}/2\,,
\label{eq:StandardClosedUniverse-t}
\eeqa
\end{subequations}
with $a_\text{max}=10$ from \eqref{eq:BCSstandarduniverse}.
The time-symmetric solution \eqref{eq:StandardClosedUniverse-a} has
Big Bang coordinate time $t_\text{BB}=-\pi a_\text{max}/2$ and
Big Crunch coordinate time $t_\text{BC}=+\pi a_\text{max}/2$.
For $t \downarrow t_\text{BB}$, the behavior of $a(t)$ approaches that of the
flat ($k=0$) FRW universe, $a(t)\propto (t-t_\text{BB})^{2/3}$.

These results for a standard closed FRW universe without vacuum energy
serve as benchmark for those of the nonstandard universe discussed in
Secs.~\ref{sec:closed-FRW-universe-vacuum-dynamics} and \ref{sec:nonstandard-universe}.
For example, the comparison of the three top-row panels
in Figs.~\ref{fig:ClosedFRWgammaAnsatz} and \ref{fig:ClosedFRWstandard}
highlights the different behavior at the stationary point $t=0$.
Similarly, the time dependence or time independence
of $\rho_\text{M}\,a^3$ in the respective middle bottom-row panel indicates
the presence or absence of energy exchange between vacuum and matter.

\end{appendix}



\vspace*{0mm}
\begin{thebibliography}{99}
\bibitem{Einstein1917}
\vspace*{-2mm}
A. Einstein,
``Kosmologische Betrachtungen zur allgemeinen Relativit\"{a}tstheorie,''
Sitzungsber. Preuss. Akad. Wiss. 1917,
Phys.--Math. Klasse, p. 142;
reprinted in:  \emph{The Collected Papers of Albert Einstein,
Vol. 6, The Berlin Years: Writings 1914--1917}, edited by A. Kox et al.,
(Princeton University Press, Princeton, 1996), Doc. 43;
translated as:
``Cosmological considerations on the general theory of relativity,''
in: \emph{The Principle of Relativity},
edited by H.A. Lorentz {\it et al.}
(Dover Publ., New York, 1952), Chap. IX.

\bibitem{Weinberg1972}
S. Weinberg,
\emph{Gravitation and Cosmology}
(Wiley, New York, 1972).

\bibitem{HawkingEllis1973}
S.W. Hawking and G.F.R. Ellis,
\emph{The Large Scale Structure of Space-Time}
(Cambridge University Press, Cambridge, England, 1973).

\bibitem{Mukhanov2005}
V. Mukhanov,
\emph{Physical Foundations of Cosmology}
(Cambridge University Press, Cambridge, England, 2005).

\bibitem{Weinberg2008}
S. Weinberg,
\emph{Cosmology}
(Oxford University Press, Oxford, England, 2008).

\bibitem{KlinkhamerVolovik2008-PRD77}
F.R. Klinkhamer and G.E. Volovik,
``Self-tuning vacuum variable and cosmological constant,''
Phys. Rev. D {\bf 77}, 085015 (2008), arXiv:0711.3170.

\bibitem{KosteleckyMewes2002}
A. Kosteleck\'{y} and M. Mewes,
``Signals for Lorentz violation in electrodynamics,''
Phys. Rev. D {\bf 66}, 056005 (2002), arXiv:hep-ph/0205211.

\bibitem{KlinkhamerRisse2007}
(a) F.R. Klinkhamer and M. Risse,
``Ultrahigh-energy cosmic-ray bounds on nonbirefringent modified-Maxwell theory,''
Phys. Rev. D {\bf 77}, 016002 (2008), arXiv:0709.2502;
(b) F.R. Klinkhamer and M. Risse,
``Addendum: Ultrahigh-energy cosmic-ray bounds on nonbirefringent
modified-Maxwell theory,''
Phys. Rev. D {\bf 77}, 117901 (2008), arXiv:0709.2502;
(c) F.R. Klinkhamer and M. Schreck,
``New two-sided bound on the isotropic Lorentz-violating parameter of
modified Maxwell theory,''
Phys. Rev.  D {\bf 78}, 085026 (2008),  arXiv:0809.3217.

\bibitem{DuffNieuwenhuizen1980}
M.J. Duff and P. van Nieuwenhuizen,
``Quantum inequivalence of different field representations,''
Phys.\ Lett.\  B {\bf 94}, 179 (1980).

\bibitem{Aurilia-etal1980}
A. Aurilia, H. Nicolai, and P.K. Townsend,
``Hidden constants: The theta parameter of QCD and the cosmological constant of N=8 supergravity,''
Nucl.\ Phys.\  B {\bf 176}, 509 (1980).

\bibitem{Volovik2005}
G.E. Volovik,
``Cosmological constant and vacuum energy,''
Ann. Phys. (Leipzig)  {\bf 14}, 165 (2005), arXiv:gr-qc/0405012.

\bibitem{Volovik2003}
G.E. Volovik,
``Evolution of cosmological constant in effective gravity,''
 JETP Lett.\  {\bf 77}, 339 (2003),
arXiv:gr-qc/0302069.

\bibitem{Barcelo2007}
C. Barcelo,
``Cosmology as a search for overall equilibrium,''
JETP Lett. {\bf 84}, 635 (2007),
arXiv:gr-qc/0611090.

\bibitem{Amendola2007}
L. Amendola, G. Camargo Campos, and R. Rosenfeld,
``Consequences of dark matter--dark energy interaction on cosmological parameters
  derived from SNIa data,''
Phys. Rev.  D {\bf 75}, 083506 (2007), arXiv:astro-ph/0610806.

\bibitem{Penrose1979}
R. Penrose,
``Singularities and time-asymmetry,''
in: \emph{General Relativity: An Einstein Centenary Survey},
edited by S.W. Haw\-king and W. Israel
(Cambridge University Press, Cambridge, 1979), Chap. 12.

\bibitem{Klinkhamer2002}
F.R. Klinkhamer,
``Fundamental time asymmetry from nontrivial space topology,''
Phys. Rev. D {\bf 66}, 047701 (2002), arXiv:gr-qc/0111090.

\bibitem{Vilenkin1984-1999}
(a) A. Vilenkin,
``Quantum creation of universes,''
Phys.\ Rev.\  D {\bf 30}, 509 (1984);
(b) A. Vilenkin,
``Boundary conditions in quantum cosmology,''
Phys.\ Rev.\  D {\bf 33}, 3560 (1986);
(c) A. Vilenkin,
``The quantum cosmology debate,''
arXiv:gr-qc/9812027.

\bibitem{RatraPeebles1988}
B. Ratra and P.J.E. Peebles,
``Cosmological consequences of a rolling homogeneous scalar field,''
Phys. Rev.  D {\bf 37}, 3406 (1988).

\bibitem{Eisenstein2005}
D.J. Eisenstein {\it et al.}  [SDSS Collaboration],
``Detection of the baryon acoustic peak in the large-scale correlation function of SDSS luminous red galaxies,''
Astrophys.\ J.\  {\bf 633}, 560 (2005), arXiv:astro-ph/0501171.

\bibitem{Astier2006}
P. Astier {\it et al.}  [SNLS Collaboration],
``The Supernova Legacy Survey: Measurement of $\Omega_\text{M}$, $\Omega_\Lambda$ and $w$ from the first year data set,''
Astron. Astrophys.  {\bf 447}, 31 (2006), arXiv:astro-ph/0510447.

\bibitem{Riess2007}
A.G. Riess {\it et al.},
``New Hubble Space Telescope discoveries of type Ia supernovae at $z >1$: Narrowing constraints on the early behavior of dark energy,''
Astrophys. J. {\bf 659}, 98 (2007), astro-ph/0611572.

\bibitem{Komatsu2008}
E. Komatsu {\it et al.},
``Five-year Wilkinson Microwave Anisotropy Probe (WMAP) observations: Cosmological interpretation,''
arXiv:0803.0547v1.

\bibitem{Freedman2001}
W.L. Freedman {\it et al.} [Hubble Space Telescope Collaboration],
``Final results from the Hubble Space Telescope Key Project to measure the Hubble constant,''
Astrophys.\ J.\  {\bf 553}, 47 (2001), arXiv:astro-ph/0012376.

\bibitem{SahniStarobinsky2006}
V. Sahni and A.A. Starobinsky,
``Reconstructing dark energy,''
Int. J. Mod. Phys.  D {\bf 15}, 2105 (2006), arXiv:astro-ph/0610026.

\bibitem{KlinkhamerVolovik2008-PRD78}
F.R. Klinkhamer and G.E. Volovik,
``Dynamic vacuum variable and equilibrium approach in cosmology,''
Phys. Rev. D {\bf 78}, 063528 (2008),
arXiv:0806.2805.

\bibitem{KlinkhamerVolovik2008-JETPL}
F.R. Klinkhamer and G.E. Volovik,
``$f(R)$ cosmology from $q$--theory,''
JETP Lett. {\bf 88}, 289 (2008), arXiv:0807.3896.

\end{thebibliography}
\end{document}